\documentclass[twocolumn]{article}
\usepackage{graphicx} 
\usepackage{authblk}
\usepackage{caption}
\usepackage{booktabs}
\usepackage{multirow}
\usepackage{longtable}
\usepackage{geometry}
\usepackage{lscape}
\usepackage{hyperref}
\usepackage{amsmath, amssymb}
\usepackage[natbib=true,style=numeric-comp,backend=bibtex,useprefix=true]{biblatex}
\title{ITGPT: A Transformer Based Architecture for the Generation of Dance Dance Revolution and In the Groove Charts}
\author{Miguel O'Malley}
\affil{Max Planck Institute for Mathematics in the Sciences\\ ScaDS.AI  \\ \texttt{miguel.omalley@mis.mpg.de}}

\begin{document}

\twocolumn[{%
\renewcommand\twocolumn[1][]{#1}%
\maketitle
\begin{center}
    \centering
    \captionsetup{type=figure}
    \includegraphics[width=.9\textwidth]{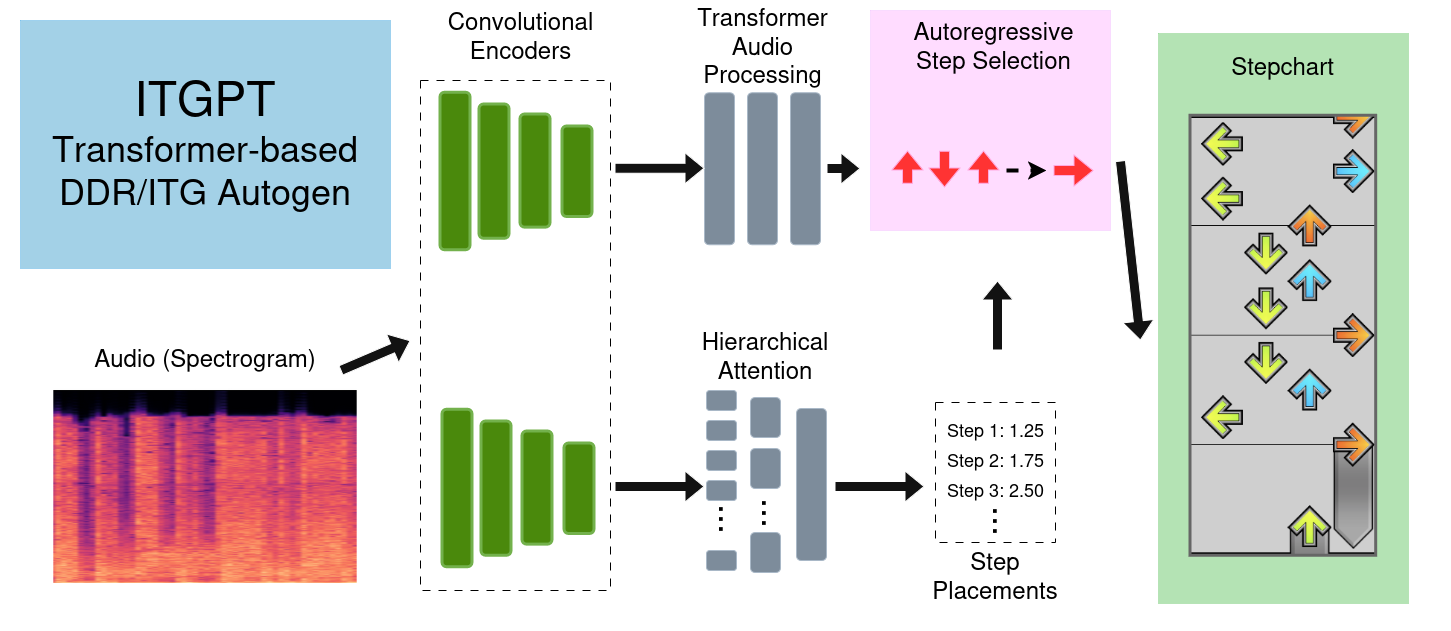}
    \captionof{figure}{The ITGPT pipeline. Audio to 4-panel chart generation.}
\end{center}%
}]

\begin{abstract}
    \textit{Dance Dance Revolution} and \textit{In the Groove} are rhythm games consisting of songs and accompanying choreography, referred to as charts. Players press arrows on a device referred to as a `dance pad' in time with steps determined by the song's chart. The process of manual chart generation is timestaking and difficult, motivating interest in automation. We propose ITGPT, a new transformer based architecture for the generation of DDR/ITG charts, and demonstrate significant improvements to generation accuracy and computational cost in comparison to predecessor work.
\end{abstract}

\section{Introduction}
\textit{Dance Dance Revolution} (DDR), and its modern counterpart \textit{In the Groove} (ITG), are rhythm games consisting of songs and charts. Players hit one of four arrows (up, down, left, right) as directed by a chart written to be played in time with an accompanying musical track. Player performance is determined by step accuracy, with steps closer to the assigned timing receiving better evaluations. A single chart may have multiple variations to accommodate difficulty settings determined by the stepcharter, and are associated with a coarser verbal difficulty level (Beginner, Easy, Medium, Hard, Expert) as well as a finer integer difficulty assignment. While verbal difficulties are expected to correlate appropriately for each song's charts, they do not necessarily correlate across distinct songs. That is, the `Hard' difficulty for one song may be more difficult than the `Medium' difficulty for another. The fine integer difficulty is expected to be objective. For example, charts rated '10' should be harder than charts rated '8'.

Charts and music are expected to correspond in a rhythmically, musically, and thematically fitting manner. While not a firm rule, steps should generally be placed where musical activity is most prominent. Depending on the desired difficulty, speed of the song (BPM), and overall styling of the music, more or fewer steps, or more or fewer technical steps may be appropriate to place. Such demands constitute a rich and challenging task for the application of modern machine learning (ML) methods.

In 2017, the authors of DDC~\cite{DDC} introduced a bipartite model structure for step placement and selection, using a hybrid convolutional/LSTM (CNN-LSTM) architecture. This model has been referred to as state-of-the-art as recently as 2023~\cite{LoveLive}. In 2025, we proposed DDCL~\cite{DDCL}, a similarly bipartite model employing $\Delta$-beat based audio modulation and ConvLSTM encoding. DDCL improved dramatically on generation quality compared to DDC, but did not include any transformer based processing.

Here we introduce ITGPT, a major departure from both DDC and DDCL. We eschew the use of LSTM in favor of a transformer based architecture. While CNN-LSTM and ConvLSTM architectures are effective at encoding local musical context, moving to a transformer based architecture allows us to better encode global context and integrate features such as difficulty and BPM. We maintain the shift to $\Delta$-beat based sampling for the step placement process, as well as the integration of audio features into the step selection process. We introduce an encoder which takes a much larger context window at a maximum of $2000$ beats by default, more than the length of most songs. We maintain the integration of fine difficulty and BPM in the step selection process. BPM detection is handled through the same ArrowVortex~\cite{ArrowVortex} algorithm used by human stepcharters, and difficulty is provided at the time of training and generation. We introduce a diagnostic model during the training process for the step placement model, which enforces alignment with BPM and difficulty parameters. Following DDC and DDCL, the ITGPT step selection model operates autoregressively, using each generated step as input for the next. We implement a much larger context window by default at 500 steps, larger than DDC (64 steps) and DDCL (256 steps). We improve generation times over DDCL by around $7\times$ by preprocessing the full audio stream for step selection before generation. We introduce a ramping multi-step training curriculum similar to~\cite{ProphetNet} in order to better structure the step selection latent state for pattern based predictions.

The ITGPT step placement model outputs 48 evenly distributed steps per beat, while our step selection model outputs 256 possible step combinations for each placed step. We evaluate our model's performance on an expanded version of the Fraxtil dataset used in the original DDC paper and DDCL papers, adding new charts created by Fraxtil which were not included in the original dataset. While DDCL was a significant improvement over the original work in DDC, ITGPT is a further improvement over DDCL still, providing significant improvement in the step placement and selection processes. 

\section{Related Work}

Dance Dance Convolution (DDC)~\cite{DDC} introduced the CNN-LSTM pipeline and the bipartite generation process for rhythm game charting. Our work in DDCL~\cite{DDCL} introduced beat modulation and ConvLSTM to this pipeline. Similar approaches have been undertaken for the game Taiko no Tatsujin (TaikoNation~\cite{TaikoNation}), osu!~\cite{OSU}~\cite{AutoOsu}~\cite{TCP}, BeatMania (GenerationMania~\cite{BeatMania}), and LoveLive~\cite{LoveLive}. An approach extending DDC to generate lower difficulties can be found in DDG~\cite{DDG} through creating a `gradation' of generation by restricting the training set to particular coarse difficulties. Dancing Monkeys~\cite{DancingMonkeys} provides a purely rule based approach to chart generation. The authors of \cite{RhythmTransformer} propose a transformer based architecture for the generation of general rhythm game charts, but the architecture appears to be limited to step placement for DDR/ITG charts.

In this work, we will explicitly compare our results to the original DDC architecture, as well as DDCL. We expect the transfomer based approach in ITGPT will significantly improve global cohesion across charts, and improve pattern recognition and reconstruction for step selection. We will also compare ITGPT against the rhythm game transformer from~\cite{RhythmTransformer}, referred to as GOCT, using their osu! four panel fine-tuned version. This is the highest performing model from their paper. DDG is another possible comparison, but in effect this would be a comparison against DDC trained on a subset of the training data. From the DDG paper, it is unclear which model should be considered as a baseline, so we elect to compare against the DDC architecture trained on the full dataset.

\section{Data}

We utilize an expanded Fraxtil dataset from the original DDC~\cite{DDC} and DDCL papers, consisting of 8 packs with 253 songs and 952 charts from the single eponymous author. This utilization of the Fraxtil dataset is meant to maintain continuity with prior work in DDC and DDCL, while expanding the available training data to improve model generalization. The expanded Fraxtil dataset contains roughly $3\times$ the content of the original version from the DDC and DDCL papers. In comparison to DDC, this dataset is larger than the original Fraxtil and ITG datasets from that work combined. Before mirroring, this gives us 584644 steps across 952 charts of information to train our step placement model with.

We maintain the up-down left-right mirroring process used in the DDC paper. In effect, this process multiplies the volume of step data produced by a given collection of charts by 4, providing our symbolic model with an effective 2338576 steps of information across 3808 charts after mirroring. The technicality of a step is unchanged by mirroring all steps left to right or up to down, as patterns are maintained by this transformation~\cite{ITGwiki}.

\begin{table}[!h]
    \centering
    \caption{The expanded Fraxtil dataset used in this paper.}
    \label{tab:dataset}
    \resizebox{\columnwidth}{!}{
        \begin{tabular}{@{}c@{}}
            \begin{tabular}{|l|r|r|}
            \hline
            \textbf{Dataset} & \textbf{Fraxtil} & \textbf{Fraxtil Expanded} \\
            \hline
            \textbf{Songs} & 90 & 253 \\
            \hline
            \textbf{Charts} & 450 & 952 \\
            \hline
            \textbf{Total Steps} & $\sim$175000 & 584644\\
            \hline
            \textbf{Total Chart Len.} & 15.3 hrs & 38.85 hrs \\
            \hline
            \end{tabular}
        \end{tabular}%
    }
\end{table}

We represent input data in two ways for our distinct step placement and selection models. For the step placement process, each chart is represented as a collection of beats, with each beat represented by the tuple (audio features, auxiliary features). The audio features term consists of a (32, 80, 3) sample of 80 melbands at 32 evenly placed timesteps across the beat. The auxiliary features term contains two constants relevant to the charting process, (BPM, integer (fine) difficulty). The target and model outputs per beat are represented by $48$ length binary vectors, representing the distribution of steps across each beat.

\begin{figure}[!tbp]
    \includegraphics[width = \columnwidth]{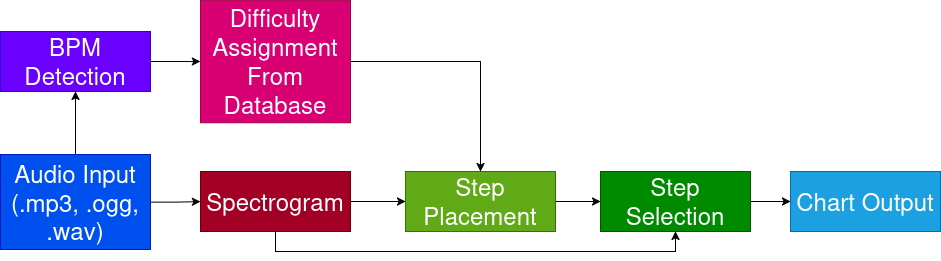}
    \caption{The ITGPT pipeline. Our pipeline is structurally identical to that established in DDCL.}
\end{figure}

\section{Model description}

Following DDC, we implement a two step process for chart generation. We first determine step placements, then step selections for each step. From DDCL, we maintain the computation of BPM as a pre-generation alignment step using the algorithm described by van de Wetering~\cite{ArrowVortexPaper} and applied in the charting software ArrowVortex~\cite{ArrowVortex}. ArrowVortex is a commonly used and ubiquitously recommended editor for stepcharting (see~\cite{ITC}). It is worth note that since our algorithm uses the same method for BPM detection as a human stepcharter, the BPM assigned to songs at generation time will be identical to that which would have been assigned for the source material. Thus, there is little to no risk of BPM misalignment between the training and generation processes.

The ITGPT step placement model takes as input audio features sampled across each beat, together with BPM and difficulty information. The goal of the step placement model is to encode each audio feature while taking BPM and difficulty as context, then produce step placements across each beat. Each output is represented as a length 48 binary vector, with 1 representing a placed step and 0 representing an absent step. In the interest of computational efficiency, ITGPT employs a one-shot structure, computing all step placements for a song or a song chunk at once. On average, ITGPT takes $0.06$s to generate a chart (see Table~\ref{tab:gen_times_onset}).

The ITGPT step selection model takes as input the full context of previously placed steps, up to 500 observations, together with audio features sampled in a given radius around each placed step, and auxiliary features indicating the timing between each step. As there are 4 arrows and 4 possible placements for each arrow (no step, step, hold, release), there are $4^4=256$ possible variations for each step. Each direction (left, right, up, down) can either be represented by a 0, 1, 2, or 3, with $0$ representing no step, $1$ representing a placed step, $2$ representing the start of a hold (where the player is expected to keep their foot on the given step until a release appears) and $3$ representing the release from a hold. Many charts have more complicated placements (e.g, ``mines'' which punish the player for stepping on them, and ``rolls'' which require the player to hit them repeatedly until a release step is observed) but in the interest of computational complexity we restrict ourselves to $4$ possible arrow selections.

\section{Methods}

\subsection{Generation}

We extract audio features from a given file and determine BPM as described above. We feed spectrogram audio features together with BPM and desired difficulty information through our step placement model to determine step placement times. Difficulties for each generated chart are determined through distribution across a census of the training set by BPM and song length. By default, 5 charts will be generated at coarse difficulties Expert, Hard, Medium, Easy, and Beginner with evenly distributed difficulties between the top and bottom 10\% for each BPM/Song length bracket. The desired difficulties may also be set by the user. We pass these step placements together with audio feature information through our step selection model to determine steps for each chart. Step selections are made after a nucleus sample with a threshold of $.9$, discarding low likelihood possibilities once a $90\%$ nucleus has been reached.

To improve generation quality, we introduce a scaling repetition penalty, similar to comparable techniques in natural language processing (NLP) but adapted for DDR/ITG. For a given step, a context window of the previous 20 placed steps is maintained. If a new step will cause a repetition of an $n$-gram of length $l\in[4,8]\subset\mathbb{N}$, we apply a scaling penalty of $p = 1.07^{l-3}$ to the offending logits before nucleus sampling. This is intended to push the model towards more varied patterns and away from lengthy repetitions.

\subsection{Audio features}

We maintain the DDC audio pipeline. This process involves ingesting each audio file as a monolithic representation taken as an average of stereo PCM audio. We then maintain three channels for each observation at the input step, representing short-time Fourier transforms (STFT) with windows of 23ms, 46ms, and 92ms, with a stride of 10ms. This methodology mirrors that used in DDC and DDCL, in turn following methodology from~\cite{BandWindows}. We reduce this representation to an 80 band mel-spectrogram representation, with log-scaling to better represent the range of human auditory perception. We organize this audio information across each beat by taking $32$ evenly spaced samples. Duplicate samples are possible if the window for a beat does not contain 32 distinct samples. Due to our larger context window, we provide the full song or 2000 beats of audio to the model for each pass, whichever is shorter.

The final inputs to our step placement model are a tensor of shape $(T, F = 80, W = 32, C = 3)$, where T represents the number of beats in a given song, along with two integers representing BPM and difficulty. For songs with variable BPM, we separate the song into segments by BPM where the BPM is consistent for at least 50 beats. Where this is impossible, we group short segments of variable BPM into segments of at least 50 beats and assign the average of the BPM values observed. We normalize each song to have 0-mean and unit standard deviation across each frequency and channel.

\begin{figure}[!tbp]\label{plot:ITGPTstepplacement}
    \includegraphics[width = \columnwidth]{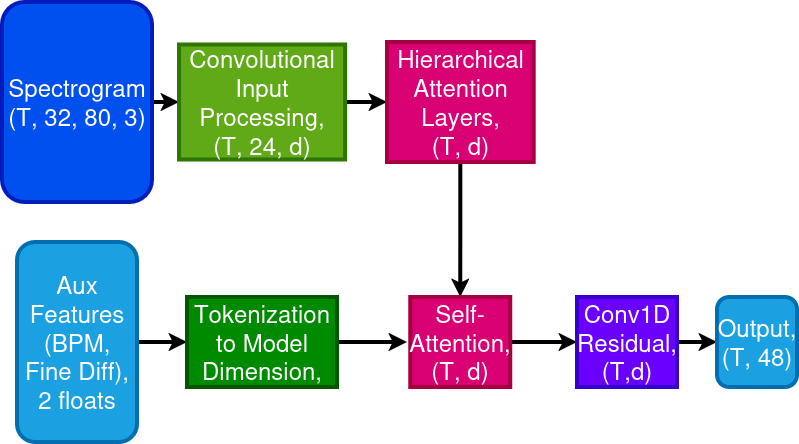}
    \caption{The ITGPT step placement (onset) model. Convolutional audio encoding is further processed through expanding self-attention layers, before combination with aux features (BPM/Difficulty), self-attention across the full song, then minor Conv1D processing for output smoothing.}
\end{figure}

\subsection{Step placement model}

The ITGPT step placement model architecture consists of a hierarchical transformer encoder coupled with specialized downstream prediction heads and multi-scale local refiners. The overall framework is regularized via a joint optimization protocol utilizing two auxiliary diagnostic networks.

\subsubsection{Audio Feature Extraction}
The input tensor representing the full spectrogram audio of a song or chunk is denoted as $\mathbf{X} \in \mathbb{R}^{T \times F \times W \times C}$ where $T$ is the total number of beats, $F = 80$ represents frequencies, $W = 32$ represents the frames sampled per beat, and $C = 3$ represents channels. For each beat, the local slice at time $t$, denoted $\mathbf{X}_t$, is processed by a convolutional encoder consisting of four convolutional layers with GELU activation. The layers are:

\begin{enumerate}
    \item \textbf{Layer 1:} (7,3) Conv2D layer with no padding and stride 1.
    \item \textbf{Layer 2:} (3,3) Conv2D layer with padding along the $W$ axis and stride 3 along the frequency axis. This reduces the frequencies by a factor of 3, similar to the maxpooling used in DDC.
    \item \textbf{Layer 3:} (3,3) Conv2D layer with no padding and stride 1.
    \item \textbf{Layer 4:} (3,3) Conv2D layer with padding along the $W$ axis and stride 3 along the frequency axis. Identical to layer 2.
\end{enumerate}
\begin{figure}[!tbp]\label{plot:ITGPTonsetencoder}
    \includegraphics[width = \columnwidth]{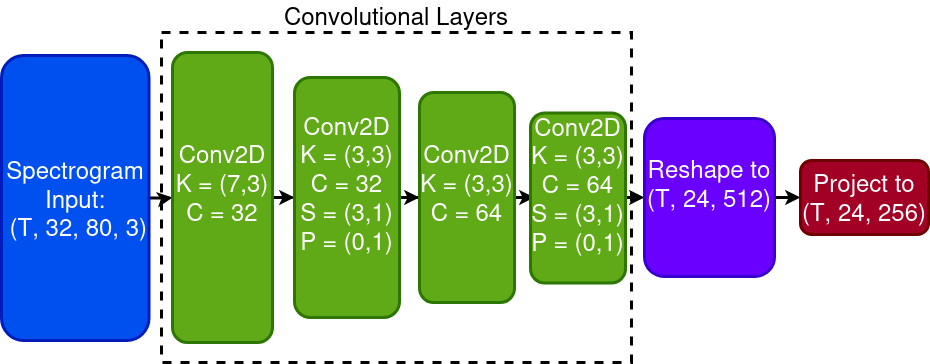}
    \caption{The ITGPT onset convolutional encoder. 32 frame per beat representations are compressed to 24 feature vectors along each beat. The hierarchical stack will reduce these further before global self-attention.}
\end{figure}

This stack extracts local feature maps of shape $T \times 64 \times 24 \times 8$. These maps are flattened across the channel and spatial frequency dimensions into a shape of $T \times 24 \times 512$, which are subsequently projected to match the core model dimension $d_{\text{model}} = 256$, yielding the base beat sequence representation $\mathbf{H}_0 \in \mathbb{R}^{T \times 24 \times d_{\text{model}}}$.

Simultaneously, the continuous conditioning variables for the beats-per-minute ($\text{BPM}$) and chart difficulty ($\text{Diff}$) are mapped into a normalized bounded range $[0, 1]$ based on predefined dataset configuration bounds (set arbitrarily at [0,50] and [30, 500]). BPM is first bucketed within 10 BPM bands, to suppress minor variations. These normalized scalars are projected through distinct MLP layers to construct context tokens $\mathbf{E}_{\text{diff}}, \mathbf{E}_{\text{bpm}} \in \mathbb{R}^{1 \times d_{\text{model}}}$.

\subsubsection{Hierarchical Framework}

To analyze the composition of the chart across multiple time horizons, we process step selection through a hierarchical stack of encoder blocks before processing across the sequence. Let $C_i$ denote the number of beats per chunk at layer $l$, and $F_i$ represent the frame resolution per beat. Then our hierarchical layers are:
\begin{enumerate}
    \item \textbf{Level 1 ($l=1$):} Processes individual beats ($C_1 = 1$, $F_{\text{pb}} = 24$).
    \item \textbf{Level 2 ($l=2$):} Processes 4-beat bars ($C_2 = 4$, $F_{\text{pb}} = 24$).
    \item \textbf{Level 3 ($l=3$):} Processes 4-bar phrases ($C_3 = 16$, $F_{\text{pb}} = 24$).
    \item \textbf{Level 4 ($l=4$):} Processes 8-bar phrases ($C_4 = 32$, $F_{\text{pb}} = 24$).
\end{enumerate}
Following Level 4, we project each beat's worth of 24 frames to a compressed 6 frame representation. This lower-resolution representation is subsequently processed at 

\begin{enumerate}

    \item \textbf{Level 5} ($C_5 = 64$, $F_{\text{pb}} = 6$).
    
\end{enumerate}

\begin{figure}[!tbp]\label{plot:ITGPThierarchy}
    \includegraphics[width = \columnwidth]{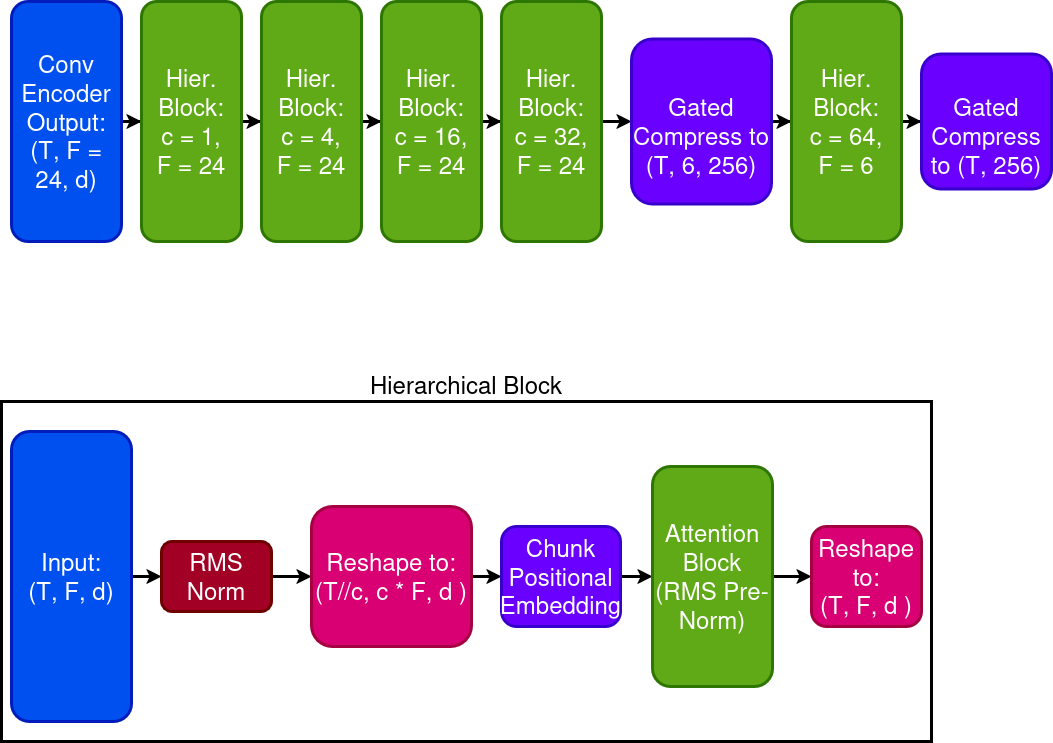}
    \caption{The ITGPT step placement hierarchical processing. Expanding receptive fields allows the model to first contextualize local audio signals, before explicitly addressing larger structural chunks.}
\end{figure}

\subsubsection{Global self-attention and convolution}
At \textbf{Level 6}, the $6$ compressed frames are projected into a single representation vector. We inject a global learned positional embedding $\mathbf{P}$,  re-append aux token embeddings, then process the sequence through a stack of transformer layers. Before output, we pass the model through a final Conv1D pass to smooth local inconsistencies before output to density and onset prediction heads.

\subsubsection{Step Placement Diagnostic Network}

To preserve difficulty and BPM through generation, we pre-train a diagnostic network to estimate auxiliary parameter values from model outputs. Specifically, we train a diagnostic network to estimate BPM when provided difficulty and a chart, and difficulty when provided BPM and the chart. We first RMS normalize, project, and embed the BPM and difficulty information, as in the main model. We then process the placed steps for a given chart through a Conv1D layer with a size 7 filter, before adding the embedded auxiliary value tensor to the chart representation. We then process the resulting time-series tensor through a bidirectional LSTM before applying mean-pooling and a final MLP pass before output. The diagnostic network's loss, computed using model inputs and outputs, is added to the total loss before optimization.

\begin{figure}[!tbp]\label{plot:diag_model}
    \includegraphics[width = \columnwidth]{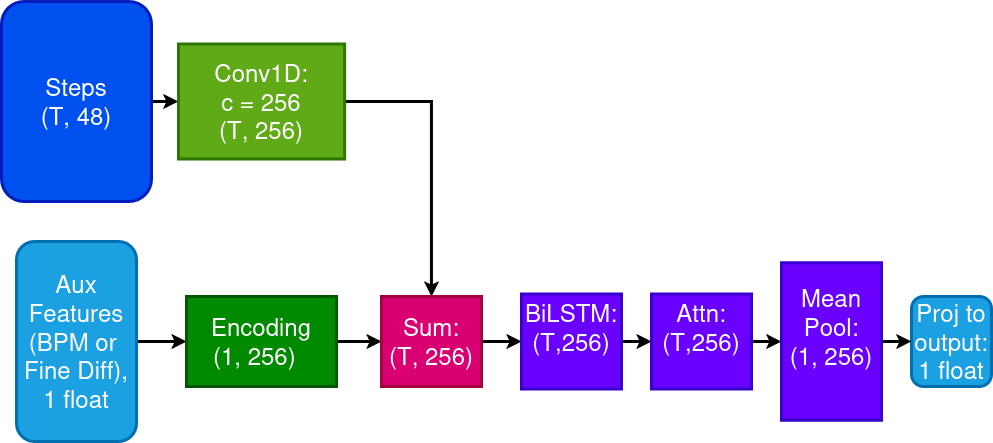}
    \caption{The ITGPT diagnostic model. Inputs are placed steps, per the step placement model.}
\end{figure}

\subsubsection{Training methodology}

We optimize using AdamW~\cite{AdamW} with an initial learning rate of $1e-4$ and scheduled reduction on plateau with a factor of $0.5$ and 5 epochs of patience, predicated on the validation set's loss. For loss, we employ binary crossentropy with weighting $w$ corresponding to positions in the $48$ vector beat, based on common position prioritization. Positions $[0, 12, 24, 36]$ corresponding to the downbeat, offbeat, and 16th notes are given weight 2, positions $[8, 12, 32, 40]$ corresponding to triplets and 24th notes are given weight $1.5$, and other positions are given weight $0.5$. We employ this reweighting scheme as most positions within a beat will be uniformly empty, thus a minor reweighting scheme is appropriate to address class imbalance. We further implement early stopping with a patience of 10 epochs, predicated on the validation set's PR-AUC. Loss via the diagnostic network is evaluated by the MSE of the target and predicted cumulative probability sums, to increase penalization for substantially different BPM and diff predictions. The diagnostic networks are trained on ground truth and their loss calculations are added to the main model's per chunk loss with a lambda of $\lambda_{diag} = 0.005$. Thus, for prediction $y_p$, target $y_t$, BPM $B$, and fine integer difficulty $D$, our full loss computation is given by

\[ \mathcal{L}(y_t, y_p, B, D) = w\cdot\text{BCE}(y_t, y_p) + \lambda_{diag}\mathcal{L}_\text{Diag}   \]

where

\begin{align}
    \mathcal{L}_\text{Diag} = \,&\mathcal{L}_{cum}(B, \text{Diag}_{B}(y_p, D)) \ + \nonumber \\
    &\mathcal{L}_{cum}(D, \text{Diag}_{D}(y_p, B))
\end{align}

\begin{figure}[!tbp]\label{plot:DDCLstepselection}
    \includegraphics[width = \columnwidth]{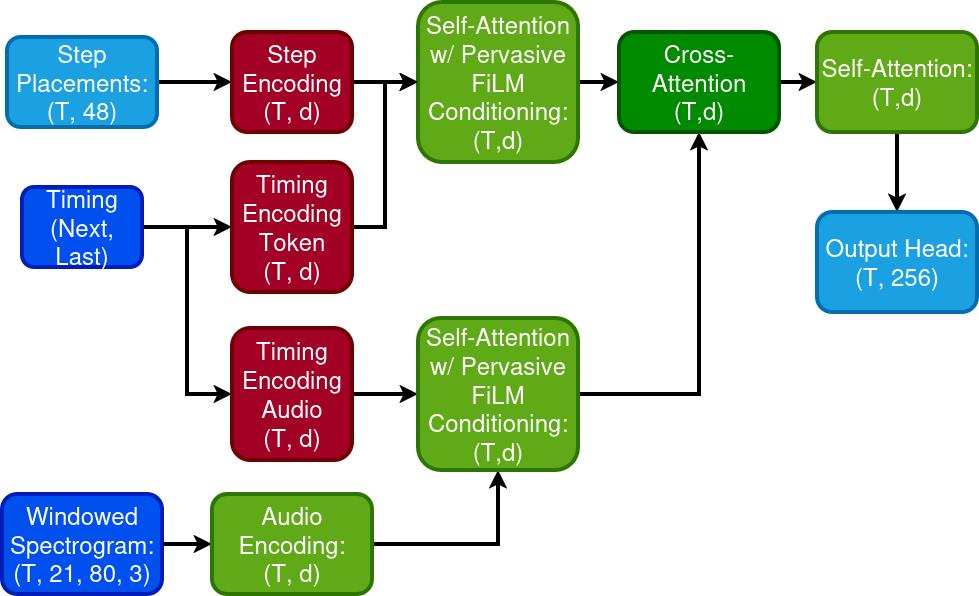}
    \caption{The ITGPT step selection model.}
\end{figure}

\subsection{Step selection}

We approach the process of step selection as an autoregressive sequence generation task. As a departure from predecessor work, we now consider the full song's audio features as input, or the last 500 steps, whichever is shorter. Each step is paired with 21 frames of audio input, representing 10 frames of past and future information for 200ms of total audio context. 

\subsubsection{Step selection audio encoder}

Input audio is formed as a 4-dimensional tensor $\mathbf{X}_{\text{audio}} \in \mathbb{R}^{T \times S \times F \times C}$, where $T$ represents sequence length, $S$ denotes the frame dimension, $F$ denotes frequency bins, and $C$ represents channels. This representation is processed by a convolutional frontend consisting of four layers with GELU activation. The four convolutional layers are identical to those used in the step placement convolutional encoder, with the exception of the final layer which produces twice the channels, for better audio resolution. Following the convolutional encoder, we collapse the audio representation along the frequency and time within beat axes, producing a representation of shape $(T, 104, d_{model}/2)$.

We apply residual vector quantization to better generalize across our dataset. Residual vector quantization in audio encoding is a highly standardized practice intended to contextualize and discretize audio representations for more comprehensive encoding (see~\cite{encodec},\cite{DAC},\cite{soundstream}). In particular, we implement the RVQ from~\cite{soundstream}, found \href{https://github.com/lucidrains/vector-quantize-pytorch}{here}. We employ 4 codebooks with $\dim=d_{model}/2$, using cosine similarity to evaluate codebook and commitment losses, and a dead code threshold of $2$. Codebook and commitment losses are passed through the model and evaluated as a sum with the total loss. 

After RVQ, we add two positional embeddings for the frequency and frame axes, reshaped to match the quantized representation, then pass through a transformer encoder layer. This layer is intended to better extract and contextualize local audio relationships highlighted by the convolutional layers. We apply attention pooling across the pooled frame/frequency dimensions to produce a $128$-dim feature vector for each placed step. We project these vectors to a single $d_{model}$ feature vector per-step for the main model.

\begin{figure}[!tbp]\label{plot:ITGPTsymencoder}
    \includegraphics[width = \columnwidth]{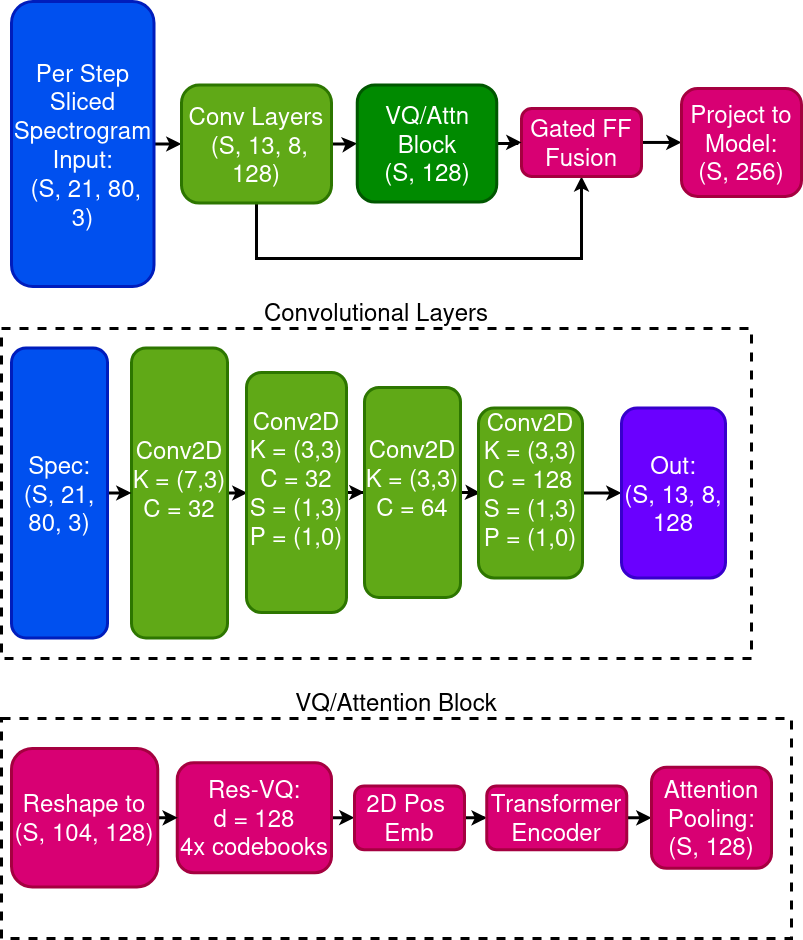}
    \caption{The ITGPT step selection audio encoder.}
\end{figure}

\subsubsection{Step selection model}

Along with the encoded and compressed audio signal of shape $(T, d_{model})$, the ITGPT step selection model takes as input a tensor of shape $(T, 2)$ representing the $\Delta$-beat time between the last step and the next step, respectively. The model iterates autoregressively through placed steps, taking as input the last 500 steps, or the full song if fewer than 500 steps have been placed.

For a step with $S$ previously placed steps, the step history tensor has shape $(S, 256)$, representing the 256 valid step classes. The aux tensor is encoded through two embedding layers, to a tensor of shape $(T, d_{model})$ which will be referred to as the audio aux tensor, and a tensor of shape $(S, d_{model})$ which will be referred to as the step aux tensor. The step aux tensor is sliced to match the number of steps in memory. We employ pervasive FiLM conditioning before every self attention block with each respective aux tensor. The audio tensor is passed through two self-attention blocks with FiLM conditioning applied before each attention pass. We then perform gated fusion between the audio and aux tensors to form a single audio context tensor of shape $(T, d_{model})$.

The step tensor of shape $(S, 256)$ is passed through a linear layer to project to $(S, d_{model})$, and then passed through six self-attention blocks with FiLM conditioning applied before each attention pass. Two cross attention blocks are then applied with the step tensor as the query and the audio context tensor as the key and value. We then apply six more self-attention blocks before output through a linear layer to a head of shape $(S+1, 256)$ representing the logits for each step class. We then apply softmax to produce a probability distribution over the 256 possible step classes.

\subsubsection{Training methodology}

Our step selection model is trained using AdamW~\cite{AdamW} with an initial learning rate of $1e-4$ and scheduled reduction on plateau with a factor of $0.5$ and 5 epochs of patience, predicated on the validation set's loss. 

During training, we have the network instead output 4 split heads, representing predictions for the current step as well as the following $3$ steps. This process is intended to improve the model's ability to learn patterns across multiple steps, inspired by a similar technique from ProphetNet~\cite{ProphetNet}. At generation time, the additional predictions are not used. During training, we apply scaling dropout on the step history, scaling up to $.2$. That is, eventually $20\%$ of the step history is blanked during training to aid in generalization. The warmup period consists of $100000$ charts passed through the model, during which the dropout is scaled linearly.

Loss is calculated as the weighted sum of categorical cross entropy over the $4$ predicted steps. Step $0$ received weight 1. For each step $i\in\{1,2,3\}$, we compute a step exponent:

\[ \lambda_i = 1 + (\frac{i}{4}).  \]

For a training run of $N$ total steps and a current step of $s$, we compute the weight for step $i$ as
\[ w_i(s,N) = (s/N)^{\lambda_i}. \]

This weighting scheme causes the model to focus on earlier steps earlier on in the training procedure, with later predictions receiving increased weight further into training. Close to the end of training, $s/N$ becomes close to $1$, and so the model begins to weight all predictions nearly evenly. The final loss is then computed as the sum of the cross entropy for each step, weighted by the step weights, along with the VQ loss from the residual vector quantization.

\section{Experiments}

We apply the same 8/1/1 split used in the original DDC and DDCL papers for the expanded Fraxtil dataset. We further maintain the chart difficulty segregation implemented in DDC, as charts for the same song at different difficulty levels often present with similar or identical sections. That is, all difficulties for any given chart will appear in the same split.

All testing was performed on a system using a single Nvidia RTX 3080 with 10GB of VRAM.

\subsection{Step placement evaluation}

We compare the ITGPT step placement model against two alternate versions, intended to highlight the value of choices in the training and design methodologies. The ITGPT (NH) or no hierarchy version of the model omits the hierarchical processing, in favor of a more aggressive downsampler before passing to global attention. This is a much more traditional version of an encoder only step placement model structure. The ITGPT (ND) or no diagnostic version of the model omits the diagnostic network, and is trained solely on the binary crossentropy loss for step placements. We also compare our models to DDC and DDCL as predecessor works, as well as the pre-trained model from \cite{RhythmTransformer}, denoted GOCT, which we test using the checkpoints the authors provide. Following conventions from that paper, GOCT is compared to the threshold maximized versions of each model as GOCT is a token based classifier and cannot be optimized further in this way. DDC is benchmarked using the same $20$ms accuracy window methodology used in the original paper, and in the $20$ms version of DDC from the DDCL paper. 

\begin{figure}[!ht]\label{plot:GenComp}
    \includegraphics[width = \columnwidth]{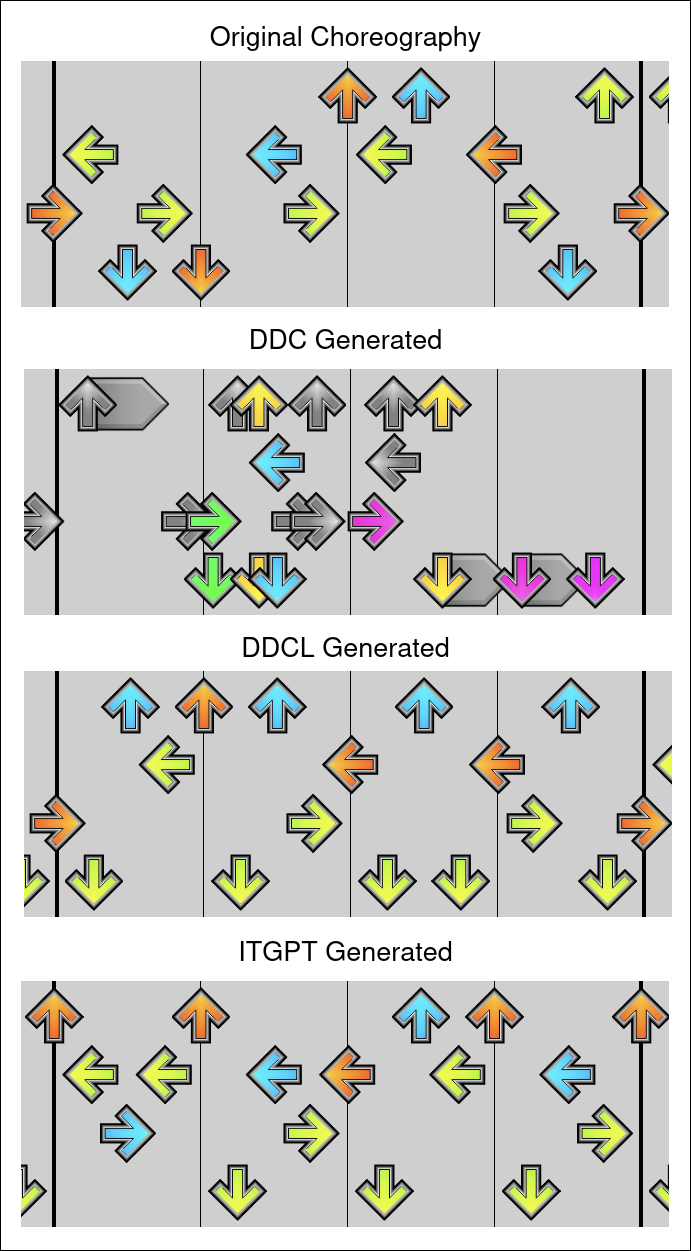}
    \caption{A comparison of DDC, DDCL, and ITGPT chart generation outputs for a snippet from the song \textit{Bad Ketchup} by Ladyscraper. The color of each arrow represents the timing placement of the arrow (red represents down beats, blue offbeats, light green 16th notes, etc.) DDCL and ITGPT are both marked improvements over DDC, but ITGPT exhibits far more similar patterning to the original work.}
\end{figure}

\subsubsection{Metrics}

We report mean performance across all predictions for F1score, precision, and recall with a flat threshold set at .5 in Table~\ref{tab:all_metrics}. Following methodology from DDC and DDCL, we also report metrics with threshold optimization for F1score. To better comprehend performance for full charts, we also report these metrics averaged across charts instead of by prediction. Metrics averaged this way are denoted with the `cht' subscript. Precision-recall AUC (PR-AUC) and loss (BCE) are reported with chart averaging. For a more granular breakdown of model performance at different difficulty levels, we report model performance averaged across each coarse (verbal) and fine (integer) difficulty level in tables~\ref{tab:all_coarse_metrics} and~\ref{tab:all_fine_metrics}, respectively. GOCT metrics are not reported for coarse difficulties as their architecture discards this identifier at evaluation time.

We observe the greatest benefit from threshold tuning from the DDC model, with a smaller but still significant benefit for DDCL. ITGPT shows the smallest improvement, which appears to be a result of the diagnostic model. Threshold tuning still provides some improvement, but not to such an extent that tuning is required for acceptable ITGPT performance. The ITGPT (ND) model, trained without the diagnostic model, exhibits a similar level of performance improvement from threshold tuning to that of DDCL, implying that one benefit of the diagnostic model is the elimination of the need to tune for thresholds in generation. This is an especially plausible conclusion, since the original generation method used by DDC relied on threshold selection to generate different difficulties.

One of the main benefits exhibited by the DDCL step placement model was a dramatic improvement among lower difficulty charts. ITGPT extends this trend, with the greatest improvement observed among charts at the Beginner difficulty level, and significant but more modest improvements across the board. The shift to a transformer with access to global context appears to be a dramatic improvement over both the DDC and DDCL models, as ITGPT is superior to DDCL in almost every metric, for every difficulty.

GOCT is fairly close to ITGPT if no threshold tuning is performed. However, as GOCT is a token based classifier model, the most likely token is selected before decoding, and thus there is no threshold to optimize to improve performance.

\begin{table*}[!ht]
    \centering
    \caption{Onset model performance across various metrics. Max modifiers indicate metrics calculated with thresholds optimized for F1score. Chart averaging indicates metrics averaged across charts instead of across placed steps.}
    \resizebox{2\columnwidth}{!}{
        \begin{tabular}{|l|c|c|c|c|c|c|}
            \hline
            \textbf{Metric} & \textbf{DDC} & \textbf{DDCL} & \textbf{GOCT} & \textbf{ITGPT} & \textbf{ITGPT (NH)} & \textbf{ITGPT (ND)} \\ \hline
            F1Score & 0.5006 & 0.7033 & - & \textbf{0.7801} & 0.7490 & 0.7701 \\ \hline
            Prec. & 0.7030 & 0.7239 & - & 0.7538 & \textbf{0.7740} & 0.7454 \\ \hline
            Recall & 0.3887 & 0.6838 & - & \textbf{0.8084} & 0.7255 & 0.7964 \\ \hline
            Max F1Score & 0.7317 & 0.7598 & 0.7754 & \textbf{0.8022} & 0.7914 & 0.7998 \\ \hline
            Max Prec. & 0.6606 & 0.6995 & 0.7500 & \textbf{0.7621} & 0.7427 & 0.7545 \\ \hline
            Max Recall & 0.8198 & 0.8313 & 0.8027 & 0.8467 & 0.8469 & \textbf{0.8509} \\ \hline
            F1Score\textsubscript{\textbf{cht}} & 0.4850 & 0.6543 & - & 0.7348 & 0.7039 & \textbf{0.7352} \\ \hline
            Prec\textsubscript{\textbf{cht}} & 0.6721 & 0.6818 & - & 0.7282 & \textbf{0.7394} & 0.7270 \\ \hline
            Recall\textsubscript{\textbf{cht}} & 0.4636 & 0.6848 & - & 0.7604 & 0.6935 & \textbf{0.7640} \\ \hline
            Max F1\textsubscript{\textbf{cht}} & 0.6540 & 0.7046 & 0.7539 & 0.7709 & 0.7549 & \textbf{0.7711} \\ \hline
            Max Prec.\textsubscript{\textbf{cht}} & 0.5960 & 0.6483 & \textbf{0.7434} & 0.7324 & 0.7069 & 0.7280 \\ \hline
            Max Rec.\textsubscript{\textbf{cht}} & 0.7855 & 0.8092 & 0.7539 & 0.8300 & \textbf{0.8376} & 0.8356 \\ \hline
            CE\textsubscript{\textbf{cht}} & 0.0913 & 0.0583 & - & 0.0374 & \textbf{0.0369} & 0.0372 \\ \hline
            AUC\textsubscript{\textbf{cht}} & 0.6356 & 0.6990 & - & 0.8030 & 0.7758 & \textbf{0.8035} \\ \hline
            \end{tabular}
    }            
    \label{tab:all_metrics}
\end{table*}

\subsubsection{Generation times}

Here we make brief note that while most models evaluated have negligible generation times for step placement due to one-shot architectures, the model used in~\cite{RhythmTransformer} is autoregressive and thus has significantly increased computational cost compared to other models shown here. While the increased cost is not so severe that GOCT cannot be used, ITGPT is both more accurate and nearly $50\times$ faster.

\begin{table}[!ht]
    \centering
    \caption{Generation time comparison between ITGPT and GOCT. We omit DDC and DDCL as both are one-shot models and thus comparable to ITGPT.}
    \resizebox{\columnwidth}{!}{
        \begin{tabular}{|l|c|c|}
            \hline
            \textbf{Model} & \textbf{ITGPT} & \textbf{GOCT} \\ \hline
            Num charts & 72 & 72 \\ \hline
            Total time (s) & 4.45 & 215.28 \\ \hline
            Avg per cht (s) & 0.06 & 2.99 \\ \hline
            \end{tabular}
    }            
    \label{tab:gen_times_onset}
\end{table}

\begin{table*}[ht]
    \centering
    \caption{Average symbolic model performance by prediction (top) and macro chart-level (bottom). Symbolic model performance across all models, averaged by prediction (top) and chart (bottom). We note that ITGPT exhibits the best performance across all metrics, with the exception of held note accuracy, where DDCL performs slightly better.}
    \label{tab:sym_avg}
    \resizebox{2\columnwidth}{!}{
        \begin{tabular}{lcccccc}
        \toprule
        \textbf{Model} & \textbf{Loss} & \textbf{Acc} & \textbf{Top 2 Acc} & \textbf{Top 3 Acc} & \textbf{HoldAcc} & \textbf{StepAcc} \\
        \midrule
        \textbf{ITGPT} & 1.9850 & \textbf{0.5908} & \textbf{0.8259} & \textbf{0.9033} & 0.3782 & \textbf{0.6204} \\
        \textbf{ITGPTs}     & 2.0109 & 0.5813          & 0.8165          & 0.8965          & 0.3482 & 0.6138          \\
        \textbf{DDCL}      & 1.1924 & 0.5533          & 0.8100          & 0.9001          & \textbf{0.4115} & 0.5683          \\
        \textbf{DDC}       & 1.5366 & 0.4523          & 0.7135          & 0.8330          & 0.1891 & 0.4840          \\
        \bottomrule
        \end{tabular}
    }
    
    \vspace{1.5em}
    
    \resizebox{2\columnwidth}{!}{
        \begin{tabular}{lcccccc}
        \toprule
        \textbf{Model} & \textbf{Loss} & \textbf{Accuracy} & \textbf{Top 2 Acc} & \textbf{Top 3 Acc} & \textbf{HoldAcc} & \textbf{StepAcc}\\
        \midrule
        \textbf{ITGPT} & 1.9343 & \textbf{0.6058} & \textbf{0.8337} & \textbf{0.9097} & 0.3778          & \textbf{0.6249} \\
        \textbf{ITGPTs}     & 1.9544 & 0.6005          & 0.8268          & 0.9047          & 0.3690          & 0.6222          \\
        \textbf{DDCL}      & 1.1701 & 0.5689          & 0.8173          & 0.9042          & \textbf{0.4067} & 0.5809          \\
        \textbf{DDC}       & 1.5807 & 0.4298          & 0.6901          & 0.8279          & 0.1943          & 0.4553          \\
        \bottomrule
        \end{tabular}
    }
\end{table*}

\begin{figure*}[!ht]\label{plot:topk_acc}
    \includegraphics[width = 2\columnwidth]{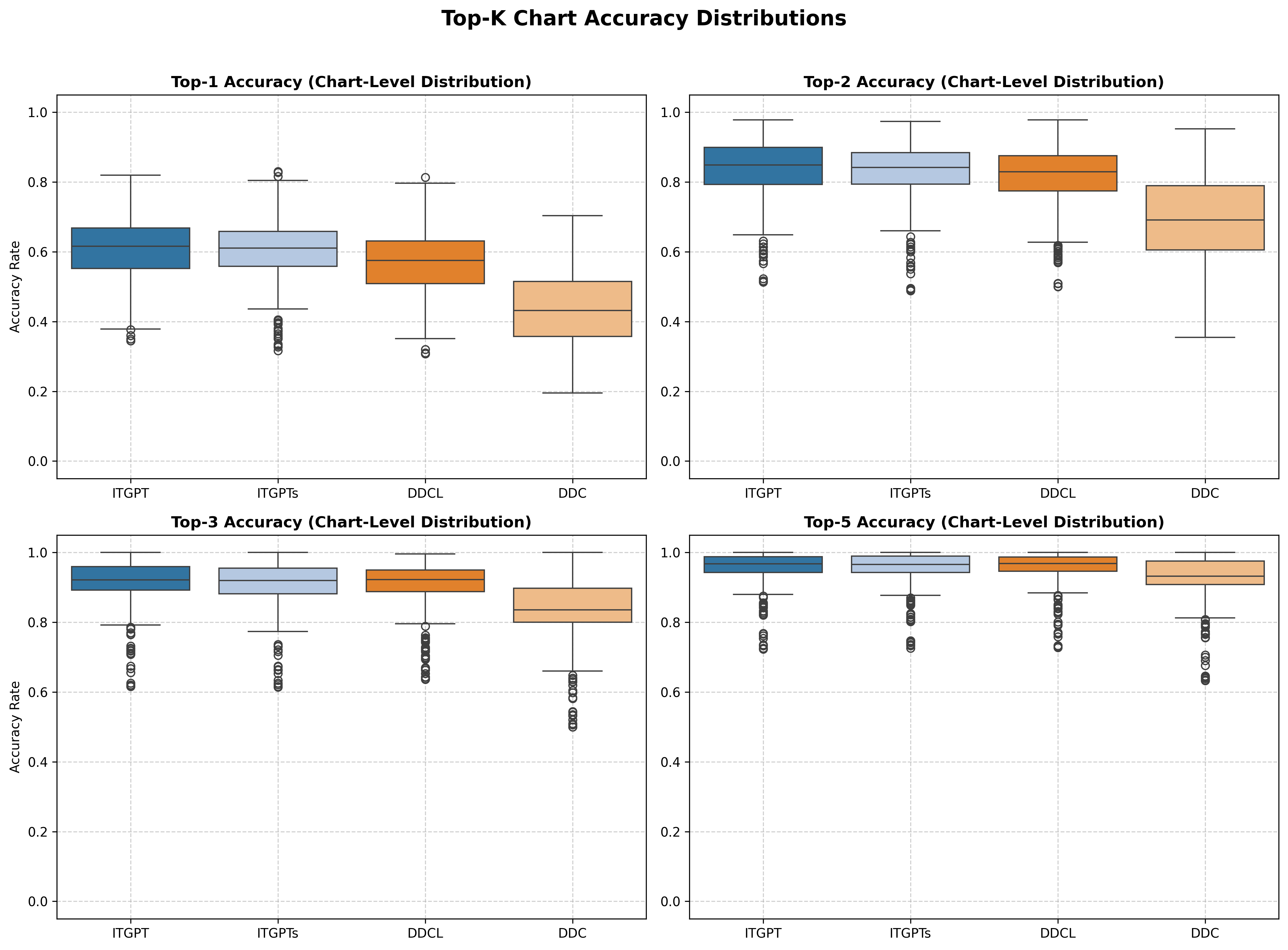}
    \caption{Box plots representing top-k accuracies for each model. We note that while ITGPT's big version exhibits the best performance, the significantly faster small model is fairly close.}
\end{figure*}
\begin{figure}[h]\label{plot:hold_perf}
    \includegraphics[width = \columnwidth]{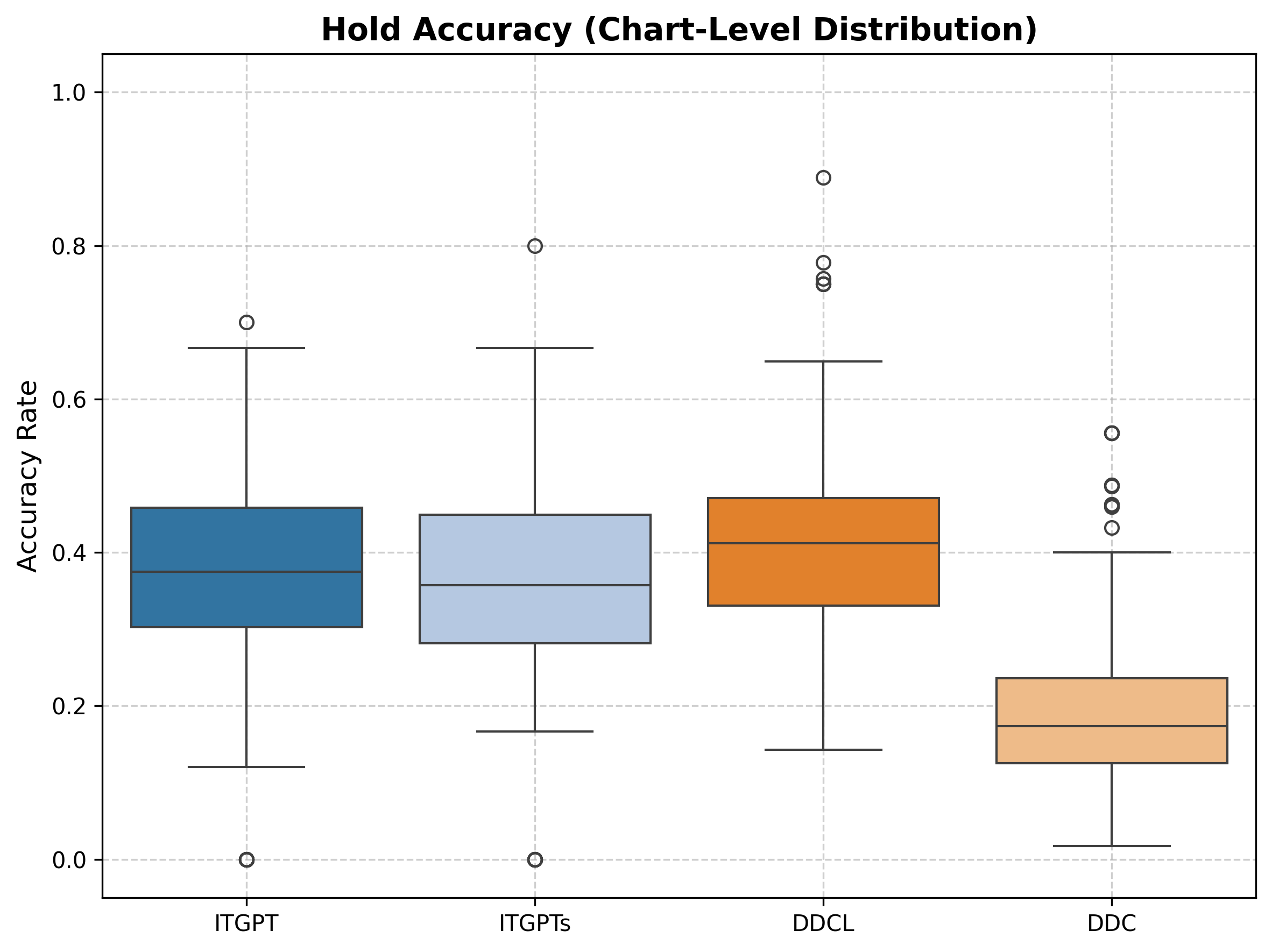}
    \caption{Box plot representing performance on held notes. Strangely, this is the single metric in which DDCL is the strongest model.}
\end{figure}

\subsection{Step selection evaluation}

We compare our ITGPT step selection model against a version of the same model with a smaller encoder, which we denote ITGPTs. Compared to the version we describe above, the smaller ITGPT version eschews the audio input level transformer in favor of a large convolutional compression block with a $(13,3)$ kernel, to reduce each step's audio context to a single feature vector. While this model is substantially faster to train, it is not meaningfully faster at generation time (see Table~\ref{tab:sym_gen_times}). We also compare our model against the DDC and DDCL step selection models. GOCT is timing only for DDR/ITG, and as such does not have the capability to select steps for ITG charts.

\subsubsection{Metrics}

We report average accuracy and loss (CCE) across all placed steps for each model. We further report top-$k$ accuracy for each model, where $k\in\{1,2,3,5\}$. In some cases, top $2$ accuracy is likely a better judge of true model accuracy than raw accuracy, since in stepcharting for DDR/ITG, some sequential step choices (left then down vs left then up for example) are genuinely a choice. We further report accuracy and loss averaged across charts instead of across placed steps. Loss as reported for ITGPT contains loss for the $4$ step prediction method we describe above, and is thus not comparable to DDC/DDCL. We also report accuracy for held notes and non-held notes separately, as held notes are a less common step type and thus more difficult to predict.

Similar to results observed from the DDCL paper, ITGPT exhibits significant improvement over DDCL and DDC, but not to the same degree exhibited by the step placement model. This is likely attributable to the fact that step selection is a more difficult and creative task, where multiple interpretations (and indeed, full mirrorings) of the same chart can be correct at the same time. Step placement, on the other hand, is much closer to onset detection and is thus a task with mostly right and wrong answers.

Strangely enough, DDCL exhibits better performance on held notes than ITGPT. While DDC remains catastrophically bad at held note prediction, DDCL is able to achieve a held note accuracy of $0.4115$ compared to ITGPT's $0.3782$. It is likely this is attributable to the ConvLSTM encoder, since a held note often corresponds with a held tone, which would be especially easy to extract across beat-separated frames through a ConvLSTM encoder.

\subsection{Generation times}

We measure average step selection generation times across the test split of our 8/1/1 expanded Fraxtil dataset. Step selection, as the autoregressive portion of the generation pipeline, is by far the most consequential factor in user experienced wait times for chart generation. To whit, in table~\ref{tab:gen_times_onset}, we observe an average generation time per chart of $0.06$ seconds. Notably, ITGPT improves on generation times compared to DDC, and is around $7\times$ faster than DDCL. Improvement over DDC is likely attributable to the architectural ability for ITGPT to process convolutional input once per song, then iterate purely over tokens for the remaining generation steps. This allows ITGPT to leverage a faster transformer based architecture in purely symbolic generation. In turn, DDC is much faster than DDCL since no audio input is processed whatsoever. 

\begin{table}[!ht]
    \centering
    \caption{Generation times for each model tested. We note ITGPT presents greater performance improvements than DDCL over DDC, while also improving generation speed. While ITGPTs is about twice as fast to train, it has no advantage at generation time over ITGPT.}
    \resizebox{\columnwidth}{!}{
        \begin{tabular}{lcccccc}
            \toprule
            \textbf{Model} & \textbf{ITGPT} & \textbf{ITGPTs} & \textbf{DDCL} & \textbf{DDC} \\
            \midrule
            \textbf{Num charts} & 296 & 296 & 296 & 296 \\
            \textbf{Avg. time (s)} & 4.3866 & 4.3361 & 31.0900  & 6.2737  \\
            \textbf{Max time (s)} & 18.4468 & 18.1742  & 130.3949  & 26.5616  \\
            \textbf{Min time (s)} & 0.4382 & 0.4333  & 3.1607 & 0.6386   \\
            \bottomrule
            \end{tabular}
    }            
    \label{tab:sym_gen_times}
\end{table}

\section{Discussion}

There are a number of areas for improvement to our approach. Primarily, a much larger dataset may exhibit some performance improvement. While GOCT from~\cite{RhythmTransformer} is trained on osu! datasets, their substantially improved results using a transformer architecture with a much larger dataset ($10\times$ Fraxtil-sized osu! vs our $3x$ Fraxtil) imply that a larger DDR/ITG dataset would produce an even better model, whether ITGPT or GOCT based. However, the stronger performance of ITGPT in comparison to GOCT, even using a one-shot architecture, implies that any larger dataset should likely be DDR/ITG native and not sourced from other rhythm games. Further, if a stronger GOCT model is constructed, the computational cost advantage of ITGPT would remain. It is also possible that ITGPT could be improved by adopting the GOCT approach to sampling at a BPM determined window at the time of spectrogram generation, instead of resampling along 10ms windows. However, the performance difference between GOCT and ITGPT is fairly uniform across difficulties (which tend to correlate with higher BPM), so we expect any benefit would be limited. 

A unified step placement/selection pipeline remains a possibility, but a distant one. For step placement, each beat contains $48$ binary positions, which is a representation that could possibly be reduced to around $8$ positions if we limit the maximal complexity to $24$th notes. Already this is a $2^8=256$ class classification problem if performed per-beat, or an $8$ label multi-label binary classification problem if performed per-step. If the additional $256$ per-step classes are added, the problem balloons to at minimum a completely unreasonable $256^8$ class classification problem, or at best an $8$ label, $256$ class multilabel/multiclass classification problem. Perhaps each step within a beat could be generated autoregressively, reducing each beat to an $8$ step, $256$ class multi-label classification problem, but we suspect the effect of this would be solely to exacerbate the drift issues associated with autoregressive step placement.

\section{Conclusions}

We introduce ITGPT, a transformer based model for the automatic generation of DDR/ITG charts. ITGPT represents a significant improvement over prior works, especially for lower difficulty charts. We further introduce a new dataset of DDR/ITG charts, expanding the existing Fraxtil dataset commonly used to evaluate DDR chart performance.

We introduce a new hierarchical transformer architecture for step placement, which allows the model to consider the full chart structure when making predictions. We consider this contribution individually significant, as it is bespoke for DDR/ITG charts as a task and particularly well-suited to tasks with well-defined boundaries and temporal structure, such as beat-delineated music.

We further introduce a diagnostic model for improved consistency during training for auxiliary parameters such as BPM and difficulty. While not itself an original idea in general, we are unaware of any similar approaches taken in the pursuit of rhythm game generation, and thus consider this contribution individually novel.

We introduce a new transformer based model for step selection, dramatically improving generation times through the use of mixed processing. Audio is processed once per chart, then used throughout generation, yielding even faster times than DDC, which does not use audio at all. ITGPT is about $7\times$ faster than DDCL (see Table~\ref{tab:sym_gen_times}). We further introduce a robust generation pipeline, integrating domain knowledge from rhythm games with techniques from NLP to perform highly accurate charting.

\section{Code Availability}

All code for this paper can be found at this project's \href{https://github.com/miguelomalley/ITGPT}{github}.

\section{Funding}

We gratefully acknowledge partial funding from the Alexander von Humboldt Foundation.

\printbibliography
\appendix
\section*{Appendix}

\renewcommand{\thetable}{A\arabic{table}} 
\setcounter{table}{0} 

\onecolumn
\begin{longtable}{|c|c|r|r|r|r|r|r|}
    \caption{All Metrics by Fine (Integer) Difficulty and Model} \label{tab:all_fine_metrics} \\
    \hline
    \textbf{Metric} & \textbf{Diff.} & \textbf{DDC} & \textbf{DDCL} & \textbf{GOCT} &\textbf{ITGPT} & \textbf{ITGPT (NH)} & \textbf{ITGPT (ND)} \\
    \hline
    \endfirsthead
    
    \hline
    \textbf{Metric} & \textbf{Diff.} & \textbf{DDC} & \textbf{DDCL} & \textbf{GOCT}& \textbf{ITGPT} & \textbf{ITGPT (NH)} & \textbf{ITGPT (ND)} \\
    \hline
    \endhead
    
    \hline
    \multicolumn{7}{r}{\emph{Continued on next page}} \\
    \endfoot
    
    \hline
    \endlastfoot
    \multirow{12}{*}{F1 Score} & 1 & 0.237887 & 0.345704 & - & 0.555398 & 0.472772 & \textbf{0.586783} \\ 
    & 3 & 0.449043 & 0.702762 & - & 0.750893 & 0.743661 & \textbf{0.759200} \\ 
    & 4 & 0.463938 & 0.768015 & - & 0.812379 & 0.769064 & \textbf{0.816972} \\ 
    & 5 & 0.579838 & 0.706815 & - & 0.738970 & \textbf{0.739258} & 0.735267 \\ 
    & 6 & 0.518100 & 0.627068 & - & \textbf{0.689448} & 0.649349 & 0.685083 \\ 
    & 7 & 0.593889 & 0.734263 & - & \textbf{0.800521} & 0.722717 & 0.789994 \\ 
    & 8 & 0.599181 & 0.677430 & - & \textbf{0.741127} & 0.731411 & 0.730693 \\ 
    & 9 & 0.592145 & 0.715236 & - & 0.761263 & 0.770049 & \textbf{0.778120} \\ 
    & 10 & 0.636111 & 0.652165 & - & \textbf{0.708790} & 0.679721 & 0.691311 \\ 
    & 11 & 0.520058 & 0.718407 & - & \textbf{0.792486} & 0.768294 & 0.789467 \\ 
    & 12 & 0.503756 & 0.699257 & - & \textbf{0.785233} & 0.743668 & 0.762625 \\ 
    & 13 & 0.432123 & 0.739541 & - & 0.833206 & 0.815423 & \textbf{0.848672} \\ 
    & 14 & 0.647681 & 0.728814 & - & 0.723521 & \textbf{0.774454} & 0.718981 \\ 
    & 15 & 0.493949 & 0.813567 & - & \textbf{0.899466} & 0.798533 & 0.819563 \\ 
    & 16 & 0.403346 & 0.738972 & - & \textbf{0.744702} & 0.687992 & 0.740119 \\ 
    & 17 & 0.226109 & 0.780770 & - & \textbf{0.812284} & 0.806769 & 0.809877 \\ 
    & 18 & 0.242424 & 0.814956 & - & \textbf{0.847380} & 0.832862 & 0.843145 \\ \hline
   \multirow{12}{*}{Precision} & 1 & 0.163650 & 0.235650 & - & \textbf{0.557812} & 0.499722 & 0.555460 \\ 
    & 3 & 0.420759 & 0.690947 & - & 0.740537 & \textbf{0.752685} & 0.747020 \\ 
    & 4 & 0.391762 & 0.741403 & - & 0.759006 & \textbf{0.798124} & 0.756091 \\ 
    & 5 & 0.684539 & 0.773934 & - & 0.799971 & \textbf{0.808004} & 0.779317 \\ 
    & 6 & 0.622195 & 0.668049 & - & \textbf{0.720314} & 0.693020 & 0.719073 \\ 
    & 7 & 0.705961 & 0.740374 & - & 0.754893 & \textbf{0.797092} & 0.745695 \\ 
    & 8 & 0.739353 & 0.747329 & - & 0.764815 & 0.755373 & \textbf{0.775276} \\ 
    & 9 & \textbf{0.872102} & 0.757871 & - & 0.695594 & 0.795964 & 0.740437 \\ 
    & 10 & \textbf{0.846463} & 0.741492 & - & 0.735990 & 0.726148 & 0.705282 \\ 
    & 11 & \textbf{0.889049} & 0.817729 & - & 0.783160 & 0.803576 & 0.791515 \\ 
    & 12 & \textbf{0.917811} & 0.825824 & - & 0.819456 & 0.841109 & 0.791334 \\ 
    & 13 & \textbf{0.961836} & 0.785795 & - & 0.743627 & 0.791162 & 0.753445 \\ 
    & 14 & \textbf{0.961353} & 0.792295 & - & 0.698539 & 0.893657 & 0.678481 \\ 
    & 15 & \textbf{0.966942} & 0.886871 & - & 0.860346 & 0.916756 & 0.843164 \\ 
    & 16 & \textbf{0.977477} & 0.651418 & - & 0.642857 & 0.583472 & 0.629412 \\ 
    & 17 & \textbf{0.981481} & 0.725093 & - & 0.736471 & 0.767290 & 0.706389 \\ 
    & 18 & \textbf{0.984026} & 0.769753 & - & 0.786621 & 0.807417 & 0.752999 \\ \hline
   \multirow{12}{*}{Recall} & 1 & 0.491742 & \textbf{0.735973} & - & 0.588159 & 0.506672 & 0.655762 \\ 
    & 3 & 0.528859 & 0.724297 & - & 0.769723 & 0.759861 & \textbf{0.785098} \\ 
    & 4 & 0.584792 & 0.796610 & - & 0.876340 & 0.748504 & \textbf{0.890173} \\ 
    & 5 & 0.571809 & 0.663544 & - & 0.700101 & 0.690109 & \textbf{0.716850} \\ 
    & 6 & 0.462038 & 0.593098 & - & \textbf{0.662114} & 0.617868 & 0.659666 \\ 
    & 7 & 0.548975 & 0.729238 & - & \textbf{0.853708} & 0.669026 & 0.852518 \\ 
    & 8 & 0.518920 & 0.626054 & - & \textbf{0.725347} & 0.714795 & 0.696710 \\ 
    & 9 & 0.483754 & 0.682089 & - & \textbf{0.871599} & 0.763116 & 0.833010 \\ 
    & 10 & 0.514259 & 0.602152 & - & 0.715492 & 0.659758 & \textbf{0.723524} \\ 
    & 11 & 0.389691 & 0.643747 & - & \textbf{0.807609} & 0.743493 & 0.790215 \\ 
    & 12 & 0.350129 & 0.612806 & - & \textbf{0.757864} & 0.673628 & 0.738044 \\ 
    & 13 & 0.284282 & 0.700673 & - & 0.958594 & 0.855637 & \textbf{0.972790} \\ 
    & 14 & 0.488344 & 0.674750 & - & 0.750357 & 0.683310 & \textbf{0.764622} \\ 
    & 15 & 0.355469 & 0.774351 & - & \textbf{0.952841} & 0.727672 & 0.839237 \\ 
    & 16 & 0.254098 & 0.853717 & - & 0.884892 & 0.838129 & \textbf{0.898082} \\ 
    & 17 & 0.127772 & 0.845709 & - & 0.905497 & 0.850530 & \textbf{0.948891} \\ 
    & 18 & 0.138241 & 0.865799 & - & 0.918312 & 0.859964 & \textbf{0.957810} \\ \hline
   \multirow{12}{*}{Max F1} & 1 & 0.287090 & 0.380732 & \textbf{0.682667} & 0.629883 & 0.559773 & 0.644667 \\ 
    & 3 & 0.533472 & 0.761503 & 0.793103 & \textbf{0.806560} & 0.796996 & 0.805474 \\ 
    & 4 & 0.501203 & 0.793531 & 0.800385 & 0.827655 & \textbf{0.831598} & 0.826430 \\ 
    & 5 & 0.684367 & 0.771010 & 0.809525 & 0.804900 & 0.789063 & \textbf{0.810966} \\ 
    & 6 & 0.583051 & 0.693979 & 0.662595 & \textbf{0.729961} & 0.721041 & 0.717724 \\ 
    & 7 & 0.693088 & 0.770199 & 0.793405 & \textbf{0.807122} & 0.793111 & 0.799980 \\ 
    & 8 & 0.693500 & 0.754076 & \textbf{0.769453} & 0.769259 & 0.767219 & 0.767514 \\ 
    & 9 & 0.782738 & 0.753903 & 0.739411 & 0.795255 & \textbf{0.801234} & 0.791799 \\ 
    & 10 & \textbf{0.770411} & 0.697974 & 0.698565 & 0.737474 & 0.719038 & 0.732626 \\ 
    & 11 & \textbf{0.805321} & 0.773353 & 0.766053 & 0.804550 & 0.802085 & 0.803671 \\ 
    & 12 & 0.787782 & 0.746447 & 0.651397 & \textbf{0.792516} & 0.781313 & 0.777863 \\ 
    & 13 & 0.825932 & 0.825632 & \textbf{0.909772} & 0.850959 & 0.851054 & 0.850723 \\ 
    & 14 & \textbf{0.790609} & 0.732183 & 0.643852 & 0.744428 & 0.775444 & 0.765672 \\ 
    & 15 & 0.900865 & 0.893859 & 0.869317 & \textbf{0.916323} & 0.900216 & 0.901460 \\ 
    & 16 & 0.800469 & 0.743287 & \textbf{0.804680} & 0.793081 & 0.776353 & 0.793061 \\ 
    & 17 & 0.814761 & 0.790480 & 0.811270 & 0.814419 & 0.812116 & \textbf{0.829811} \\ 
    & 18 & 0.836700 & 0.829460 & 0.822218 & 0.849146 & 0.844535 & \textbf{0.862158} \\ \hline
   \multirow{12}{*}{Max Precision} & 1 & 0.188820 & 0.280065 & \textbf{0.719254} & 0.581493 & 0.466162 & 0.585640 \\ 
    & 3 & 0.434732 & 0.669165 & \textbf{0.744110} & 0.723780 & 0.722652 & 0.738128 \\ 
    & 4 & 0.373341 & 0.708567 & 0.736750 & 0.748589 & \textbf{0.778694} & 0.748901 \\ 
    & 5 & 0.603156 & 0.729632 & \textbf{0.852426} & 0.754716 & 0.739197 & 0.772380 \\ 
    & 6 & 0.514572 & 0.581600 & \textbf{0.720072} & 0.665915 & 0.622143 & 0.638060 \\ 
    & 7 & 0.614185 & 0.704350 & \textbf{0.795698} & 0.758033 & 0.733228 & 0.762200 \\ 
    & 8 & 0.619515 & 0.686440 & \textbf{0.754686} & 0.723949 & 0.714269 & 0.739496 \\ 
    & 9 & 0.741042 & 0.702579 & 0.712072 & 0.759257 & \textbf{0.769997} & 0.730922 \\ 
    & 10 & 0.745817 & 0.699490 & 0.677145 & \textbf{0.746383} & 0.689112 & 0.719532 \\ 
    & 11 & 0.782940 & 0.749681 & 0.735957 & \textbf{0.790521} & 0.777957 & 0.777684 \\ 
    & 12 & 0.787720 & 0.768794 & 0.637249 & \textbf{0.794447} & 0.780101 & 0.764963 \\ 
    & 13 & 0.809755 & 0.742081 & \textbf{0.881664} & 0.762544 & 0.777399 & 0.779660 \\ 
    & 14 & 0.818660 & 0.772152 & 0.610561 & 0.776744 & \textbf{0.893855} & 0.802817 \\ 
    & 15 & 0.858154 & 0.871022 & \textbf{0.898828} & 0.878462 & 0.856618 & 0.840888 \\ 
    & 16 & 0.802353 & 0.673152 & 0.711331 & 0.890882 & \textbf{0.956140} & 0.820513 \\ 
    & 17 & \textbf{0.820538} & 0.683673 & 0.756464 & 0.756198 & 0.745367 & 0.784701 \\ 
    & 18 & 0.787639 & 0.738947 & 0.761377 & 0.784113 & 0.775028 & \textbf{0.824059} \\ \hline
   \multirow{12}{*}{Max Recall} & 1 & 0.692460 & 0.692479 & 0.652741 & 0.735736 & \textbf{0.799860} & 0.762888 \\ 
    & 3 & 0.728112 & 0.894493 & 0.856804 & \textbf{0.918905} & 0.900110 & 0.898516 \\ 
    & 4 & 0.825329 & 0.906873 & 0.877195 & \textbf{0.927779} & 0.899240 & 0.924944 \\ 
    & 5 & 0.799291 & 0.832564 & 0.780044 & \textbf{0.886612} & 0.858687 & 0.878280 \\ 
    & 6 & 0.673331 & \textbf{0.863252} & 0.617705 & 0.811077 & 0.859552 & 0.832688 \\ 
    & 7 & 0.819605 & 0.850232 & 0.796443 & \textbf{0.869495} & 0.866500 & 0.848427 \\ 
    & 8 & 0.796143 & \textbf{0.842130} & 0.790495 & 0.822523 & 0.831785 & 0.812164 \\ 
    & 9 & 0.837056 & 0.820698 & 0.795748 & 0.846628 & 0.843582 & \textbf{0.868925} \\ 
    & 10 & \textbf{0.806773} & 0.707675 & 0.744154 & 0.730396 & 0.758420 & 0.753271 \\ 
    & 11 & 0.832508 & 0.802791 & 0.803312 & 0.823996 & 0.830307 & \textbf{0.834625} \\ 
    & 12 & 0.789962 & 0.730677 & 0.666202 & 0.794824 & 0.784699 & \textbf{0.794993} \\ 
    & 13 & 0.852187 & 0.936013 & 0.939881 & \textbf{0.965319} & 0.940156 & 0.936064 \\ 
    & 14 & \textbf{0.764417} & 0.696148 & 0.680982 & 0.714693 & 0.684736 & 0.731812 \\ 
    & 15 & 0.950759 & 0.918366 & 0.842061 & 0.958424 & 0.949548 & \textbf{0.972808} \\ 
    & 16 & 0.798595 & 0.829736 & \textbf{0.926230} & 0.714628 & 0.653477 & 0.767386 \\ 
    & 17 & 0.809065 & \textbf{0.936837} & 0.874638 & 0.882353 & 0.891996 & 0.880424 \\ 
    & 18 & 0.892280 & \textbf{0.945242} & 0.893627 & 0.925943 & 0.927738 & 0.903950 \\ \hline
   \multirow{12}{*}{Log Loss} & 1 & 0.040156 & 0.056015 & - & \textbf{0.010051} & 0.011491 & 0.011155 \\ 
    & 3 & 0.040839 & 0.028222 & - & \textbf{0.013501} & 0.013936 & 0.014459 \\ 
    & 4 & 0.055087 & 0.026070 & - & \textbf{0.012900} & 0.013602 & 0.015163 \\ 
    & 5 & 0.044197 & 0.034361 & - & 0.022910 & 0.024597 & \textbf{0.022086} \\ 
    & 6 & 0.076200 & 0.040685 & - & \textbf{0.026453} & 0.027121 & 0.026623 \\ 
    & 7 & 0.055395 & 0.046849 & - & \textbf{0.029857} & 0.033310 & 0.030976 \\ 
    & 8 & 0.108698 & 0.056035 & - & 0.037879 & 0.037429 & \textbf{0.037326} \\ 
    & 9 & 0.069304 & 0.054970 & - & 0.041326 & \textbf{0.037000} & 0.037177 \\ 
    & 10 & 0.094781 & 0.079349 & - & 0.059210 & \textbf{0.058074} & 0.062259 \\ 
    & 11 & 0.133758 & 0.083401 & - & 0.062768 & 0.061516 & \textbf{0.059882} \\ 
    & 12 & 0.174738 & 0.091209 & - & 0.080151 & \textbf{0.072069} & 0.075183 \\ 
    & 13 & 0.103240 & 0.071476 & - & 0.049819 & \textbf{0.043245} & 0.044292 \\ 
    & 14 & 0.231366 & 0.093221 & - & 0.069948 & \textbf{0.058678} & 0.071691 \\ 
    & 15 & 0.129009 & 0.055087 & - & \textbf{0.030376} & 0.043427 & 0.042415 \\ 
    & 16 & 0.303460 & 0.064559 & - & 0.051759 & 0.062912 & \textbf{0.050602} \\ 
    & 17 & 0.197552 & 0.064629 & - & 0.053813 & \textbf{0.049807} & 0.057156 \\ 
    & 18 & 0.215588 & 0.060988 & - & 0.048265 & \textbf{0.045353} & 0.050909 \\ \hline
   \multirow{12}{*}{AUC} & 1 & 0.167987 & 0.265695 & - & 0.616685 & 0.495998 & \textbf{0.639873} \\ 
    & 3 & 0.448963 & 0.729374 & - & \textbf{0.817546} & 0.799703 & 0.811184 \\ 
    & 4 & 0.405077 & 0.787340 & - & 0.839592 & 0.837731 & \textbf{0.856773} \\ 
    & 5 & 0.679712 & 0.816122 & - & \textbf{0.863205} & 0.847621 & 0.862913 \\ 
    & 6 & 0.536171 & 0.703209 & - & \textbf{0.784934} & 0.750228 & 0.761936 \\ 
    & 7 & 0.686522 & 0.793350 & - & 0.837070 & 0.826010 & \textbf{0.839176} \\ 
    & 8 & 0.689625 & 0.760436 & - & \textbf{0.796042} & 0.788131 & 0.790204 \\ 
    & 9 & 0.828556 & 0.789152 & - & 0.860109 & \textbf{0.864343} & 0.853315 \\ 
    & 10 & 0.783582 & 0.726866 & - & \textbf{0.800121} & 0.766839 & 0.795668 \\ 
    & 11 & 0.819602 & 0.789901 & - & \textbf{0.836700} & 0.827088 & 0.833823 \\ 
    & 12 & 0.818846 & 0.772095 & - & 0.827838 & \textbf{0.828543} & 0.825185 \\ 
    & 13 & 0.886051 & 0.806888 & - & 0.902182 & \textbf{0.906907} & 0.899765 \\ 
    & 14 & \textbf{0.849884} & 0.776514 & - & 0.809753 & 0.834044 & 0.814215 \\ 
    & 15 & 0.942583 & 0.936155 & - & \textbf{0.958958} & 0.948381 & 0.946579 \\ 
    & 16 & 0.871708 & 0.764324 & - & \textbf{0.885202} & 0.858684 & 0.877835 \\ 
    & 17 & \textbf{0.886802} & 0.800482 & - & 0.863895 & 0.878800 & 0.885223 \\ 
    & 18 & 0.899982 & 0.832402 & - & 0.888034 & 0.905382 & \textbf{0.912393} \\ \hline
    \end{longtable}

\begin{longtable}{|c|c|r|r|r|r|r|}
\caption{All Metrics by Coarse Difficulty and Model} \label{tab:all_coarse_metrics} \\
\hline
\textbf{Difficulty} & \textbf{Metric} & \textbf{DDC} & \textbf{DDCL} & \textbf{ITGPT} & \textbf{ITGPT (NH)} & \textbf{ITGPT (ND)} \\
\hline
\endfirsthead

\hline
\textbf{Difficulty} & \textbf{Metric} & \textbf{DDC} & \textbf{DDCL} & \textbf{ITGPT} & \textbf{ITGPT (NH)} & \textbf{ITGPT (ND)} \\
\hline
\endhead

\hline
\multicolumn{7}{r}{\emph{Continued on next page}} \\
\endfoot

\hline
\endlastfoot
\multirow{5}{*}{F1 Score} & Beginner & 0.237887 & 0.345704 & 0.555398 & 0.472772 & \textbf{0.586783} \\ 
 & Easy & 0.451751 & 0.714626 & 0.762072 & 0.748279 & \textbf{0.769704} \\ 
 & Medium & 0.566833 & 0.692552 & \textbf{0.742250} & 0.710226 & 0.736506 \\ 
 & Hard & 0.578538 & 0.704101 & \textbf{0.748957} & 0.743579 & 0.747109 \\ 
 & Challenge & 0.515150 & 0.714488 & \textbf{0.787855} & 0.757281 & 0.776051 \\ \hline
\multirow{5}{*}{Precision} & Beginner & 0.163650 & 0.235650 & \textbf{0.557812} & 0.499722 & 0.555460 \\ 
 & Easy & 0.415487 & 0.700121 & 0.743895 & \textbf{0.760946} & 0.748669 \\ 
 & Medium & 0.673378 & 0.735903 & 0.765953 & \textbf{0.773669} & 0.753717 \\ 
 & Hard & \textbf{0.817504} & 0.750258 & 0.725445 & 0.775813 & 0.738284 \\ 
 & Challenge & \textbf{0.911447} & 0.800326 & 0.779298 & 0.796122 & 0.772470 \\ \hline
\multirow{5}{*}{Recall} & Beginner & 0.491742 & \textbf{0.735973} & 0.588159 & 0.506672 & 0.655762 \\ 
 & Easy & 0.539029 & 0.737445 & 0.789108 & 0.757796 & \textbf{0.804203} \\ 
 & Medium & 0.535644 & 0.662248 & 0.731633 & 0.664657 & \textbf{0.738255} \\ 
 & Hard & 0.482132 & 0.672944 & \textbf{0.803545} & 0.729547 & 0.779714 \\ 
 & Challenge & 0.378642 & 0.657311 & \textbf{0.808405} & 0.736704 & 0.794683 \\ \hline
\multirow{5}{*}{Max F1 Score} & Beginner & 0.287090 & 0.380732 & 0.629883 & 0.559773 & \textbf{0.644667} \\ 
 & Easy & 0.527605 & 0.767326 & \textbf{0.810395} & 0.803287 & 0.809284 \\ 
 & Medium & 0.659114 & 0.749780 & \textbf{0.785068} & 0.771615 & 0.782540 \\ 
 & Hard & 0.736845 & 0.748589 & \textbf{0.778477} & 0.775305 & 0.774177 \\ 
 & Challenge & \textbf{0.812808} & 0.770536 & 0.803482 & 0.798252 & 0.801764 \\ \hline
\multirow{5}{*}{Max Precision} & Beginner & 0.188820 & 0.280065 & 0.581493 & 0.466162 & \textbf{0.585640} \\ 
 & Easy & 0.423570 & 0.676329 & 0.728291 & 0.732842 & \textbf{0.740086} \\ 
 & Medium & 0.582005 & 0.682364 & 0.731402 & 0.705645 & \textbf{0.732971} \\ 
 & Hard & 0.692193 & 0.702818 & \textbf{0.741992} & 0.736463 & 0.736525 \\ 
 & Challenge & 0.791676 & 0.746330 & \textbf{0.792735} & 0.781141 & 0.776131 \\ \hline
\multirow{5}{*}{Max Recall} & Beginner & 0.692460 & 0.692479 & 0.735736 & \textbf{0.799860} & 0.762888 \\ 
 & Easy & 0.745788 & 0.896744 & \textbf{0.920518} & 0.899951 & 0.903321 \\ 
 & Medium & 0.770478 & 0.845752 & \textbf{0.861343} & 0.861054 & 0.857704 \\ 
 & Hard & 0.800534 & 0.815218 & 0.824751 & 0.826672 & \textbf{0.826923} \\ 
 & Challenge & \textbf{0.839212} & 0.802496 & 0.821322 & 0.823545 & 0.833249 \\ \hline
\multirow{5}{*}{Log Loss} & Beginner & 0.040156 & 0.056015 & \textbf{0.010051} & 0.011491 & 0.011155 \\ 
 & Easy & 0.043430 & 0.027831 & \textbf{0.013392} & 0.013875 & 0.014587 \\ 
 & Medium & 0.055979 & 0.039492 & 0.025771 & 0.027661 & \textbf{0.025748} \\ 
 & Hard & 0.101036 & 0.061705 & 0.044567 & \textbf{0.041524} & 0.044247 \\ 
 & Challenge & 0.142458 & 0.078281 & 0.059893 & 0.058738 & \textbf{0.058465} \\ \hline
\multirow{5}{*}{PR AUC} & Beginner & 0.167987 & 0.265695 & 0.616685 & 0.495998 & \textbf{0.639873} \\ 
 & Easy & 0.440984 & 0.739913 & \textbf{0.821554} & 0.806617 & 0.819472 \\ 
 & Medium & 0.642422 & 0.779117 & \textbf{0.834731} & 0.815165 & 0.828900 \\ 
 & Hard & 0.762252 & 0.770730 & \textbf{0.827184} & 0.818911 & 0.820195 \\ 
 & Challenge & 0.839748 & 0.789803 & \textbf{0.846696} & 0.839686 & 0.845667 \\ \hline
\end{longtable}

\twocolumn

\end{document}